\documentclass[%
reprint,
 amsmath,amssymb,
aps,
]{revtex4-1}

\usepackage{graphicx}
\usepackage{dcolumn}
\usepackage{bm}

\begin{document}


\title{Efficient spin transport along Si $\langle$100$\rangle$ at room temperature}

\author{M. Ishikawa,$^{1,2}$\thanks{E-mail address: mizue.ishikawa@toshiba.co.jp} M. Tsukahara,$^{1}$ Y. Fujita,$^{1}$ M. Yamada,$^{1}$ Y.Saito,$^{2}$\thanks{Present address: Center for Innovative Integrated Electronic System, Tohoku University, 468-1 Aramaki-Aza-Aoba, Sendai 980-0845, Japan} T. Kimura,$^{3,4}$ and K. Hamaya$^{1,4}$\thanks{E-mail address:hamaya@ee.es.osaka-u.ac.jp}}

\affiliation{
$^{1}$Department of Systems Innovation, Graduate School of Engineering Science, Osaka University, 1-3 Machikaneyama, Toyonaka 560-8531, Japan.}
\affiliation{
$^{2}$$^{2}$Corporate $\&$ Research Development Center, Toshiba Corporation, 1 Komukai-Toshiba-cho, Kawasaki, Kanagawa 212-8582, Japan.}
\affiliation{
$^{3}$Department of Physics, Kyushu University, 744 Motooka, Fukuoka 819-0395, Japan.}
\affiliation{
$^{4}$Center for Spintronics Research Network, Graduate School of Engineering Science, Osaka University, 1-3 Machikaneyama, Toyonaka 560-8531, Japan.}

\date{\today}

\begin{abstract}
We find efficient spin transport in Si at room temperature in lateral spin-valve (LSV) devices. 
When the crystal orientation of the spin-transport channel in LSV devices is changed from $\langle$110$\rangle$, which is a conventional cleavage direction, to $\langle$100$\rangle$, 
the maximum magnitude of the spin signals is markedly enhanced. 
From the analyses based on the one-dimensional spin diffusion model, we can understand that the spin injection/detection efficiency in Si$\langle$100$\rangle$ LSVs is larger than that in Si$\langle$110$\rangle$ ones. 
We infer that, in Si-based LSV devices, the spin injection/detection efficiency is related to the crystallographic relationship between the magnetization direction of the ferromagnetic contacts and the orientation of the conduction-band valleys in Si. 
\end{abstract}


\maketitle
In the field of spintronics \cite{Igor,Hirohata,Taniyama,Yuasa}, spin-based logic devices using semiconductors have so far been proposed theoretically \cite{Tanaka,Saito_TSF,Saito_JEC,Dery_Nature}. 
To achieve these concepts, electrical spin injection, transport, and detection in semiconductors have been explored by using nonlocal magnetoresistance measurements in lateral spin-valve (LSV) devices with GaAs \cite{Lou_NatPhys,Ciorga_PRB,Uemura_APEX,Saito_APEX,Salis_PRB,Bruski_APL}, Si \cite{Jonker_APL,Suzuki_APEX,Saito_IEEE,Ishikawa_PRB,Jansen_PRAP}, Ge \cite{Zhou_PRB,Kasahara_APEX,Fujita_PRAP,Yamada_APEX}, GaN \cite{Bhattacharya_APL,Park_NC}, SiGe \cite{Naito_APEX} and so forth.  
Although almost all the studies have used single crystalline semiconductor layers as the pure-spin-current transport channels, there has still been lack of information on the influence of the crystal orientation on the spin injection, transport, and detection in semiconductors. 

To date, Li {\it et al}. have clarified the influence of the $g$-factor anisotropy in the Ge conduction band on the spin relaxation of electrons by combining a ballistic hot electron spin injection-detection technique with changing in-plane applied magnetic field directions \cite{Appelbaum_PRL111}. Unfortunately, the above study did not show the pure spin current transport and the anisotropic phenomena were observed only at low temperatures. 
Very recently, Park {\it et al}. reported the crystallographic-dependent pure spin current transport in GaN-based LSVs with nanowire channels at room temperature \cite{Park_NC}. 
They discussed the influence of the spontaneous polarization, interface-specific spin filtering, or the strength of the spin-orbit coupling on the pure spin current transport in GaN nanowires. However, the detailed mechanism is still an open question. 
Also, there is no information on the crystallographic effect on the pure spin current transport in other semiconductors.

In this letter, we experimentally find the efficient pure spin current transport in Si$\langle$100$\rangle$ LSV devices at room temperature. 
The enhancement in the spin injection/detection efficiency is related to the valley structures of the conduction band in Si.
This study experimentally shows the importance of the crystallographic relationship between the magnetization direction of the ferromagnetic contacts and the orientation of the conduction-band valleys in Si.

To explore the influence of the crystal orientation of the Si spin-transport channel, we designed two kinds of devices along $\langle$100$\rangle$ and $\langle$110$\rangle$, as shown in Figs. 1(a) and 1(b), on a phosphorous-doped ($n$ $\sim$ 1.3 $\times$ 10$^{19}$ cm$^{-3}$) (001)-SOI ($\sim$ 61 nm) layer. 
As a tunnel barrier, an MgO (1.1 nm) layer was deposited by electron-beam evaporation at 200 $^\circ$C on the SOI layer \cite{Ishikawa_APL}.
Then, a CoFe (10 nm) and Ru cap (7 nm) layers were sputtered on top of it under a base pressure better than 5.0 $\times$ 10$^{-7}$ Pa. The MgO and CoFe layers were epitaxially grown on (001)-SOI, where the (001)-textured MgO layer was grown on Si(001) owing to an insertion of a thin Mg layer into MgO/Si interface \cite{Saito_AIP}. 
From the detailed characterizations, the CoFe(001)$\langle$100$\rangle$/MgO(001)$\langle$110$\rangle$/Si(001)$\langle$110$\rangle$ heterostructures were confirmed \cite{Saito_AIP}. 
Conventional processes with electron beam lithography and Ar ion milling were used to fabricate LSV devices \cite{Ishikawa_PRB,Fujita_PRAP,Yamada_APEX}. 
Next, the Ru/CoFe/MgO contacts, FM1 and FM2, were patterned into 2.0 $\times$ 5.0 $\mu$m$^{2}$ and 0.5 $\times$ 5.0 $\mu$m$^{2}$ in sizes, respectively, and the width of the Si spin-transport channel was 7.0 $\mu$m. 
Finally, Au/Ti ohmic pads were formed for all the contacts. 
Note that there was no difference in the size of the spin-injector contact between Si$\langle$100$\rangle$ and Si$\langle$110$\rangle$ LSV devices \cite{Jansen_PRAP}. 
Furthermore, the current-voltage characteristics of the FM1 (FM2) contact in Si$\langle$100$\rangle$ LSV devices were identical with those in the Si$\langle$110$\rangle$ LSV ones. 
The resistivity and Hall mobility ($\mu_{\rm Hall}$) of the Si spin-transport channel were evaluated from Hall-effect measurements for Si$\langle$100$\rangle$ and Si$\langle$110$\rangle$ Hall-bar devices. 
\begin{figure}[t]
\begin{center}
\includegraphics[width=7.5 cm]{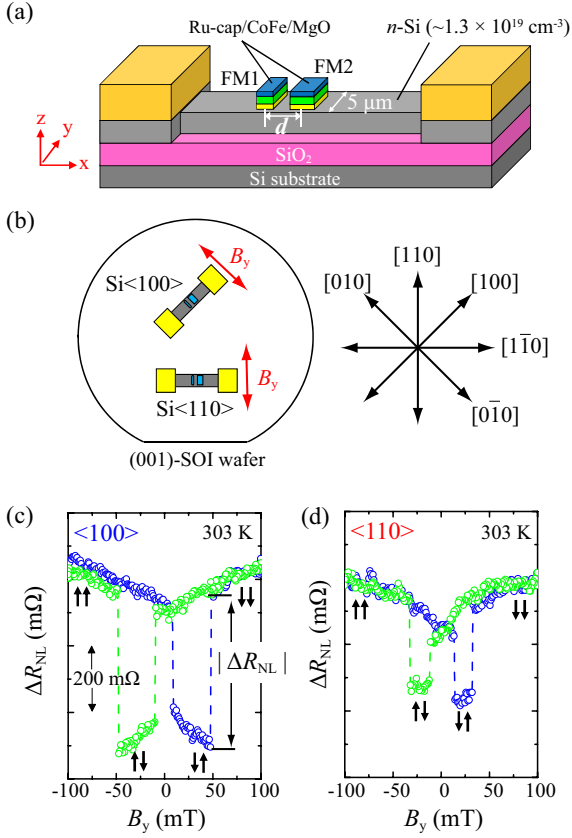}
\caption{(Color online) Schematic diagrams of (a) the lateral four-terminal device with the Si spin-transport layer and (b) the relationship between the crystal orientations, $\langle$100$\rangle$ or $\langle$110$\rangle$, and the fabricated Si spin-transport channels. (c) and (d) are  nonlocal magnetoresistance curves for Si$\langle$100$\rangle$ and Si$\langle$110$\rangle$ LSV devices ($d =$ 1.75 $\mu$m), respectively, at 303 K.}
\end{center}
\end{figure}

Figures 1(c) and 1(d) show four-terminal nonlocal magnetoresistance signals for Si$\langle$100$\rangle$ and Si$\langle$110$\rangle$ LSV devices, respectively, at a bias current ($I$) of -0.5 mA at room temperature (303 K), where the center-to-center distance ($d$) in the LSV device was 1.75 $\mu$m. 
Here in-plane external magnetic fields ($B_{\rm y}$) were applied along the directions shown in Fig. 1(b) for each Si$\langle$100$\rangle$ or Si$\langle$110$\rangle$ LSV device.
First, we can see differences in the shape and magnitude of the signals between Si$\langle$100$\rangle$ and Si$\langle$110$\rangle$ LSV devices. 
From the magnetization measurements of the epitaxial CoFe layer on MgO/Si(001), we have confirmed the presence of the magnetocrystalline easy axis along Si$\langle$110$\rangle$ (CoFe$\langle$100$\rangle$).
Namely, the magnetic fields along Si$\langle$100$\rangle$ ([100] or [010]) can contribute to the magnetization rotation because of the hard axes. 
Surely, when we applied the in-plane magnetic fields to Si$\langle$100$\rangle$ LSV devices along Si[010] for nonlocal measurements, their magnetization switching fields were larger than those of the Si$\langle$110$\rangle$ LSVs, as shown in Figs. 1(c) and 1(d). This behavior indicates that the magnetization of the narrower CoFe contact was pinned along a certain direction between Si$\langle$110$\rangle$ and the direction of $B_{\rm y}$ (Si $\langle$100$\rangle$). 
That is, although there is the shape-induced anisotropy along Si[010] in the Si$\langle$100$\rangle$ LSVs, the magnetization tends to be pinned along the magnetocrystalline easy axis along Si[110], leading to the enhancement in the magnetization switching field. In addition, the magnetization rotation of the narrower CoFe contact in the Si$\langle$100$\rangle$ LSV devices enables us to show gradual changes in the nonlocal magnetoresistance from $B_{\rm y} =$ 10 to 50 mT. 
Here we define the magnitude of spin signal, $|\Delta R_{\rm NL}|$, as the change in $\Delta R_{\rm NL}$ at the magnetization switching field from antiparallel to parallel magnetization states of the CoFe contacts. A representative $|\Delta R_{\rm NL}|$ is shown in Fig. 1(c). 
Note that $|\Delta R_{\rm NL}|$ for the Si$\langle$100$\rangle$ LSV device is nearly twice as large as that for the Si$\langle$110$\rangle$ one. 
These tendencies were observed reproductively for many LSV devices with $d =$ 1.75 $\mu$m at room temperature. 
\begin{figure}[t]
\begin{center}
\includegraphics[width= 7.5 cm]{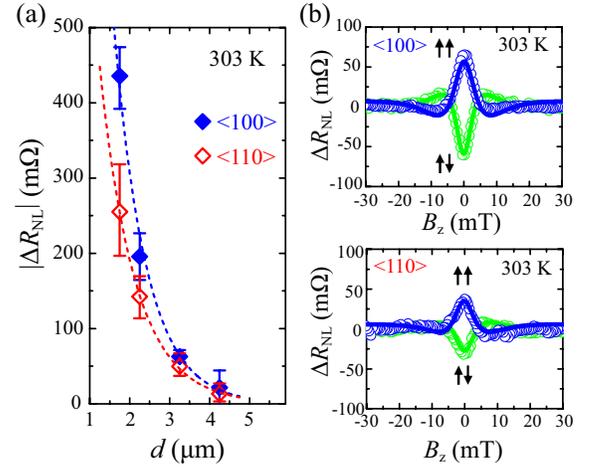}
\caption{(Color online) (a) Plots of $|\Delta R_{\rm NL}|$ versus $d$ at 303 K. The dashed line is a result fitted to Eq. (1). (b) Four-terminal nonlocal Hanle-effect curves of Si$\langle$100$\rangle$ and Si$\langle$110$\rangle$ LSV devices ($d =$ 2.25 $\mu$m) for the parallel and antiparallel magnetization states at 303 K.}
\end{center}
\end{figure}

To understand the above phenomena, we measured $d$ dependence of $|\Delta R_{\rm NL}|$ for many LSV devices, as shown in Fig. 2(a). 
Here the one data plot in Fig. 2(a) means the average of the $|\Delta R_{\rm NL}|$ value obtained from five LSV devices. 
For both Si$\langle$100$\rangle$ and Si$\langle$110$\rangle$ LSV devices, the value of $|\Delta R_{\rm NL}|$ is decreased with increasing $d$, indicating the exponential decay of $|\Delta R_{\rm NL}|$. We note that the difference between Si$\langle$100$\rangle$ and Si$\langle$110$\rangle$ becomes small in LSV devices with large $d$. 
In general,  $|\Delta$$R$$_{\rm NL}|$ in the LSVs with sufficiently large contact resistance can be expressed by the following equation: \cite{Saito_JAP,Maekawa_PRB,Fert_PRB64,Fert_PRB82}
\begin{equation}
|\Delta R_{\rm NL}| ={\frac{4|{P_{\rm inj}|}|{P_{\rm det}|}{r_{\rm Si}}{r_{\rm b}^{2}}\exp\left(-\frac{d}{\lambda_{\rm Si}}\right)} {{S_{\rm N}}\{\left(2{r_{\rm b}} + {r_{\rm Si}}\right)^{2} - {r_{\rm Si}^{2}}{\exp\left(-\frac{2d}{\lambda_{\rm Si}}\right)\}}}},
\end{equation}
where $P$$_{\rm inj}$ and $P$$_{\rm det}$ are spin polarizations of the electrons in Si created by the spin injector and detector, respectively, and ${\sqrt{|{P_{\rm inj}|}|{P_{\rm det}|}}}$ generally means the spin injection/detection efficiency of the spin injector and detector contacts. 
$r_{\rm b}$ ($\sim$ 10 k$\Omega$ $\mu$m$^{2}$) and $r_{\rm Si}$ ($=$ 0.0054 $\Omega$ cm $\times$ $\lambda_{\rm Si}$) are the spin resistances of the CoFe/MgO interface and the $n$-Si layer, respectively.  
$\lambda_{\rm Si}$ ($=$ $\sqrt{D\tau_{\rm Si}}$, where $D$ and $\tau_{\rm Si}$ are the diffusion constant and the spin lifetime, respectively) is the spin diffusion length in Si, $S_{\rm N}$ ($=$ 0.305 $\mu$m$^{2}$) is the cross-sectional area of the Si spin transport layer. Using Eq. (1), we can fit the experimental data in Fig. 2(a) and extract $\lambda_{\rm Si}$ for both Si$\langle$100$\rangle$ ($\lambda_{\rm Si} $ $\sim$ 0.80 $\mu$m) and Si$\langle$110$\rangle$ ($\lambda_{\rm Si} $ $\sim$ 0.88 $\mu$m). Also, using $D$ values of 5.03 cm$^{2}$/s for Si$\langle$100$\rangle$ ($\mu_{\rm Hall} =$ 87.0 cm$^{2}$/Vs) and 5.17 cm$^{2}$/s for Si$\langle$110$\rangle$ ($\mu_{\rm Hall} =$ 89.5 cm$^{2}$/Vs), estimated from the Hall mobility \cite{Flatte_PRL}, we can roughly estimate $\tau_{\rm Si}$ of 1.3 ns and 1.5 ns for Si$\langle$100$\rangle$ and Si$\langle$110$\rangle$, respectively. 
This implies that the difference in the spin relaxation between Si$\langle$100$\rangle$ and Si$\langle$110$\rangle$ is relatively small compared to other parameters. 
On the other hand, the obtained ${\sqrt{|{P_{\rm inj}|}|{P_{\rm det}|}}}$ value of $\sim$ 0.16 for Si$\langle$100$\rangle$ is valuably larger than that (${\sqrt{|{P_{\rm inj}|}|{P_{\rm det}|}}}$ $\sim$ 0.11) for Si $\langle$110$\rangle$. Thus, we can infer that the spin injection/detection efficiency of the contacts depends on the crystal orientation of the spin-transport channel on Si. 
\begin{table}[b]
\begin{center}
\caption{Comparison of the extracted parameters at room temperature between Si$\langle$100$\rangle$ and Si$\langle$110$\rangle$ LSVs.}
\vspace{2mm}
\begin{tabular}{l|c|c} \hline
& Si$\langle$100$\rangle$ & Si$\langle$110$\rangle$ \\ \hline
$D$ (cm$^{2}$/s) & 5.03 & 5.17 \\
${\tau_{\rm Si}}$  (ns) &  $\sim$1.4 &  $\sim$1.4 \\
$\lambda_{\rm Si}$ ($\mu$m) & $\sim$0.84 & $\sim$0.85 \\
${\sqrt{|{P_{\rm inj}|}|{P_{\rm det}|}}}$ & 0.14 & 0.10 \\ \hline
\end{tabular}
\end{center}
\end{table}

By using the nonlocal Hanle analysis \cite{Jedema_Nat}, we can also confirm whether the spin relaxation behavior between Si$\langle$100$\rangle$ and Si$\langle$110$\rangle$ changes or not. 
Figures 2(b) and 2(c) display room-temperature four-terminal nonlocal Hanle-effect curves in the parallel and antiparallel magnetization states for Si$\langle$100$\rangle$ and Si$\langle$110$\rangle$ LSV devices ($d =$ 2.25 $\mu$m), respectively. 
Here using the following one dimensional spin drift diffusion model \cite{Jedema_Nat}, we can obtain the best fit curves expressed as solid lines. 
\begin{equation}
\Delta R_{\rm NL}(B_{\rm z}) = \pm A{ {\int_0^{\infty}}{\phi(t)}{\rm cos}({\omega}_{\rm L}t){\exp\left(-\frac{t}{\tau_{\rm Si}}\right)}dt},
\end{equation}
where $A =$ ${\frac{{P_{\rm inj}}{P_{\rm det}}{\rho_{\rm Si}}D}{S}}$, $\phi(t) =$ $\frac{1}{\sqrt{4{\pi}Dt}}{\exp\left(-\frac{d^{2}}{4Dt}\right)}$, $\omega$$_{\rm L}$ (= $g$$\mu$$_{\rm B}$$B$$_{z}$/$\hbar$) is the Larmor frequency, $g$ is the electron $g$-factor ($g$ = 2) in Si, $\mu$$_{\rm B}$ is the Bohr magneton. ${\rho_{\rm Si}}$ is the resistivity of Si. 
From the fitting results, the values of ${\tau_{\rm Si}}$ were estimated to be 1.4 ns and 1.3 ns for Si$\langle$100$\rangle$ and Si$\langle$110$\rangle$, respectively, implying that there is almost no difference in the spin relaxation between Si$\langle$100$\rangle$ and Si$\langle$110$\rangle$. 
On the other hand, since the amplitude of the Hanle curve between Si$\langle$100$\rangle$ and Si$\langle$110$\rangle$ is clearly different, we can judge that the value of ${\sqrt{|{P_{\rm inj}|}|{P_{\rm det}|}}}$ for Si$\langle$100$\rangle$ is clearly larger than that for Si$\langle$110$\rangle$. 
Using the above two methods, we compare the average values of $D$, ${\tau_{\rm Si}}$, $\lambda_{\rm Si}$, and ${\sqrt{|{P_{\rm inj}|}|{P_{\rm det}|}}}$ between Si$\langle$100$\rangle$ and Si$\langle$110$\rangle$ LSV devices in Table I.
From the evaluated data in Table I, we conclude that the spin injection/detection efficiency (${\sqrt{|{P_{\rm inj}|}|{P_{\rm det}|}}}$) depends on the crystal orientation of the spin-transport layer in Si-based LSV devices. 
\begin{figure}[t]
\begin{center}
\includegraphics[width= 8.5 cm]{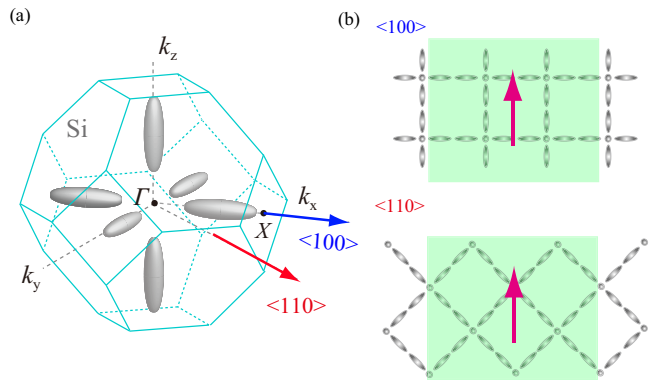}
\caption{(Color online) (a) Schematics of Brillouin zone of bulk Si. (b) Schematics of crystallographic relationship between the magnetization direction of the ferromagnetic contacts and the orientation of the conduction-band valleys in Si$\langle$100$\rangle$ (upper) and Si$\langle$110$\rangle$ (lower) LSV devices.}
\end{center}
\end{figure}

We discuss a possible origin of the large difference in the spin injection/detection efficiency between Si$\langle$100$\rangle$ and Si$\langle$110$\rangle$. 
First, the $g$-factor anisotropy in Si is negligibly small compared to that in Ge because of the weak spin orbit interaction \cite{Wilson_PR,Giorgioni_NatCom}. 
This fact means that, unlike Ge, we cannot see the change in the spin transport data only by changing the direction of the applied magnetic fields \cite{Appelbaum_PRL111}. 
Actually, we performed oblique Hanle measurements for a Si LSV device and confirmed the negligible change in the Hanle curves by changing the applied field directions (not shown here). 
Also, the interface quality of the CoFe/MgO contacts is the same between Si$\langle$100$\rangle$ and Si$\langle$110$\rangle$ LSV devices. 
Here we focus on the crystallographic orientation of the conduction band valleys in Si. 
Figure 3(a) illustrates the conduction-band valley positions in the ${\bf k}$-space in Si; six valleys are located close to the $X$ point along $\langle$100$\rangle$. 
In actual LSV devices used here [see Fig. 1(a)], since the width of the Si channel (7.0 $\mu$m) is larger than the distance ($d \le$ 4.2 $\mu$m) between the spin injector and spin detector, we should regard the pure spin current transport in Si as a two-dimensionally equivalent phenomenon even though we used two different spin-transport channel along $\langle$100$\rangle$ and $\langle$110$\rangle$. 
Thus, the anisotropic spin relaxation could not be detected between $\langle$100$\rangle$ and $\langle$110$\rangle$, as discussed in Table 1.
On the other hand, there is a difference in the configuration between the magnetization direction of the ferromagnetic contacts and the crystal orientation of the conduction-band valleys in the Si channel, as depicted in Fig. 3(b). In this situation, we can expect the difference in the spin-related electronic band structures at the CoFe/MgO/Si interface between Si$\langle$100$\rangle$ and Si$\langle$110$\rangle$
LSV devices. Thus, we speculate that the difference in the above crystallographic relationship is one of the origins of the difference in ${\sqrt{|{P_{\rm inj}|}|{P_{\rm det}|}}}$ between Si$\langle$100$\rangle$ and Si$\langle$110$\rangle$ LSV devices. 
To elucidate the detailed mechanism, further experimental and theoretical studies are required.


In summary,  we experimentally found the efficient pure spin current transport in Si$\langle$100$\rangle$ LSV devices at room temperature. 
We infer that the enhancement in the spin injection/detection efficiency is related to the crystallographic relationship between the magnetization direction of the ferromagnetic contacts and the orientation of the conduction-band valleys in Si.
This study indicates the importance of the consideration of the crystallographic orientation in the spin-transport channel even in spintronic devices.  

\vspace{5mm}
The authors acknowledge Dr. H. Sugiyama of Toshiba Corporation for useful discussion about the magnetocrystalline anisotropy of the CoFe layer on MgO/(001)SOI, respectively. 
This work was partly supported by a Grant-in-Aid for Scientific Research (A) (No. 16H02333) from the Japan Society for the Promotion of Science (JSPS), and a Grant-in-Aid for Scientific Research on Innovative Areas "Nano Spin Conversion Science" (No. 26103003) from the Ministry of Education, Culture, Sports, Science, and Technology (MEXT).






\end{document}